\documentclass[sigconf]{acmart}

\usepackage{listings}
\usepackage{placeins}
\definecolor{softblue}{RGB}{173, 216, 230}

\setcopyright{acmlicensed}
\copyrightyear{2026}
\acmYear{2026}
\setcopyright{cc}
\setcctype{by}
\acmConference[MSR '26]{23rd International Conference on Mining Software Repositories}{April 13--14, 2026}{Rio de Janeiro, Brazil}
\acmBooktitle{23rd International Conference on Mining Software Repositories (MSR '26), April 13--14, 2026, Rio de Janeiro, Brazil}
\acmDOI{10.1145/3793302.3793614}
\acmISBN{979-8-4007-2474-9/2026/04}

\begin{document}

\title{From Industry Claims to Empirical Reality: An Empirical Study of Code Review Agents in Pull Requests}

\author{Kowshik Chowdhury}
\email{kchowdh1@students.kennesaw.edu}
\affiliation{%
  \institution{Kennesaw State University}
  \city{Marietta}
  \state{Georgia}
  \country{USA}
}

\author{Dipayan Banik}
\email{dipayan5175@gmail.com}
\affiliation{%
  \institution{Quanta Technology}
  \city{Raleigh}
  \state{North Carolina}
  \country{USA}
}

\author{K M Ferdous}
\email{kferdous@students.kennesaw.edu}
\affiliation{%
 \institution{Kennesaw State University}
  \city{Marietta}
  \state{Georgia}
  \country{USA}
}

\author{Shazibul Islam Shamim}
\email{mshamim@kennesaw.edu}
\affiliation{%
  \institution{Kennesaw State University}
  \city{Marietta}
  \state{Georgia}
  \country{USA}
}

\begin{abstract}
Autonomous coding agents are generating code at an unprecedented scale, with OpenAI Codex alone creating over 400,000 pull requests (PRs) in two months. As agentic PR volumes increase, code review agents (CRAs) have become routine gatekeepers in development workflows. Industry reports claim that CRAs can manage 80\% of PRs in open source repositories without human involvement. As a result, understanding the effectiveness of CRA reviews is crucial for maintaining developmental workflows and preventing wasted effort on abandoned pull requests. However, empirical evidence on how CRA feedback quality affects PR outcomes remains limited. \textit{The goal of this paper is to help researchers and practitioners understand when and how CRAs influence PR merge success by empirically analyzing reviewer composition and the signal quality of CRA-generated comments. }From AIDev's 19,450 PRs, we analyze 3,109 unique PRs in \texttt{Commented} review state, comparing human-only versus CRA-only reviews. We examine 98 closed CRA-only PRs to assess whether low signal-to-noise ratios contribute to abandonment. CRA-only PRs achieve a 45.20\% merge rate, 23.17 percentage points lower than human-only PRs (68.37\%), with significantly higher abandonment. Our signal-to-noise analysis reveals that 60.2\% of closed CRA-only PRs fall into the 0--30\% signal range, and 12 of 13 CRAs exhibit average signal ratios below 60\%, indicating substantial noise in automated review feedback. These findings suggest that CRAs without human oversight often generate low-signal feedback associated with higher abandonment. For practitioners, our results indicate that CRAs should augment rather than replace human reviewers, and that human involvement remains critical for effective and actionable code review.
\end{abstract}



\maketitle

\vspace{-7.75pt}
\section{Introduction}

Recent software engineering platforms are witnessing a surge in autonomous coding agents. OpenAI Codex alone created over 400,000 pull requests in open-source GitHub repositories in less than two months since its release and code review agents (CRAs) such as coderabbitai have become routine parts of development workflows~\cite{Li_2025_SE3}. As these agents generate increasing volumes of code, code review remains critical for quality assurance, leading to CRAs being integrated into GitHub pull requests (PRs). Prior studies show that code review is a socio-technical process influenced by reviewer roles, build outcomes, and project activity~\cite{EcosystemPR_2024}~\cite{PREmpirical_2021}, indicating that merge decisions depend on various interactions inside the review discussion.

Software developers perceive these tools as beneficial for automating repetitive tasks and detecting issues early~\cite{Castaldi_2025}~\cite{CHASE_2025}~\cite{ACRPractice_2024}. Recent industry reports present an optimistic picture: Qodo's 2025 report, based on a survey of 609 developers, shows that 81\% of CRA users see quality improvements and 69\% see speed improvements, with 80\% of PRs receiving no human comments when CRAs are enabled~\cite{QodoReport_2025}. Similarly, Greptile blogs state that 82\% of developers use CRAs daily or weekly~\cite{GreptileReport_2025}. These industry reports and blogs suggest CRAs can handle most PRs with minimal human oversight.

However, empirical evidence reveals constraints. CRA adoption is limited by trust issues and lack of project-specific context~\cite{CHASE_2025}~\cite{LLMReview_2025}. A large-scale analysis found that human comments were addressed more often (60\%) than CRA comments (0.9\% to 19.2\%)~\cite{AICodeChanges_2025}, suggesting substantial noise in automated feedback. Meanwhile, CRAs often review code from the same provider~\cite{Li_2025_SE3}, raising concerns about closed-loop biases. Noisy CRA comments may lead to higher PR abandonment, which can create technical debt.

The gap between industry claims (80\% of PRs need no human comments~\cite{QodoReport_2025}~\cite{GreptileReport_2025}) and empirical findings (low CRA adoption~\cite{AICodeChanges_2025}) motivates investigating whether CRAs can effectively review PRs without human involvement and whether noisy feedback correlates with abandonment. Yet no prior work has systematically quantified signal-to-noise ratios in CRA reviews or linked them to PR outcomes.

\textit{The goal of this paper is to help researchers and practitioners understand when and how CRAs influence PR merge success by empirically analyzing reviewer composition and the signal quality of CRA-generated comments.} We focus on the following research questions:

\begin{itemize}
\item  \textbf{RQ1:} \textit{What are the differences in merge rates between human-only reviewed and CRA-only reviewed PRs in the commented review state?}
\item \textbf{RQ2:} \textit{How much does low signal-to-noise ratio in CRA comments contribute to lower merge rates in CRA-only reviewed PRs?}
\end{itemize}

To address these questions, we analyze the AIDev dataset~\cite{Li_2025_SE3}, which contains 19,450 PRs with review activity from HuggingFace \footnote{\url{https://huggingface.co/datasets/hao-li/AIDev/viewer/pr_review_comments}}. We extract 3,109 unique PRs and examine reviewer composition to distinguish between human-only, CRA-only, and mixed review scenarios. We measure merge and abandonment outcomes across these categories. We analyze CRA comments by classifying them using a two-tier keyword framework and computing signal-to-noise ratios to evaluate the relationship between CRA feedback quality and PR abandonment.

\section{Dataset Description}

Our analysis centers on the \textbf{PRReviewComment} table (19,450 records), which we enrich with additional information from related tables. This table provides the core data for our analysis. Each comment includes: \textbf{(1) \texttt{user}:} the name or identifier of the CRA or human who posted the comment, \textbf{(2) \texttt{user\_type}:} categorical classification identifying reviewers as ``User'' (human) or ``Bot'' (CRA), \textbf{(3) \texttt{body}:} textual content of review comments, essential for signal-to-noise ratio analysis, and \textbf{(4) \texttt{pull\_request\_url}:} API endpoint referencing the parent PR.

To enable outcome analysis, we enrich each comment with additional attributes by merging data from the \textbf{PRReview} and \textbf{PullRequest} tables. From \textbf{PRReview}, we obtain the review \texttt{state} attribute, which categorizes review feedback as \texttt{COMMENTED} (general feedback without explicit approval or rejection), \texttt{APPROVED} (approval to merge), \texttt{CHANGES\_REQUESTED} (explicit request for modifications), or \texttt{DISMISSED} (review dismissed or withdrawn). From \textbf{PullRequest}, we obtain the PR \texttt{state} (open or closed) and \texttt{merged\_at} (timestamp in ISO format, null if not merged), which are critical for determining PR outcomes. These tables are merged using \texttt{pr\_id} as the foreign key, and then joined to the comment-level data using \texttt{pull\_request\_review\_id}.

We construct our analytical dataset by aggregating enriched records by \texttt{pr\_id}, including only PRs with at least one review comment, producing 3,177 unique PRs. To ensure inclusion of only code review agents (CRAs), unique GitHub bot names were first extracted from the \texttt{user} field. These bots were then manually classified by their functionality, and entities of github bots (e.g., \texttt{github-actions[bot]}) that perform CI/CD and workflow automation rather than code review feedback were excluded. This filtering produced a final dataset of 3,109 PRs reviewed by actual CRAs. This aggregation (1) collects comment bodies into lists for signal detection, (2) aggregates review metadata to classify reviewer composition (human-only, CRA-only, or mixed), and (3) computes total comments per PR.

\section{Methodology}

For each of the 3,109 unique PRs, we examine the \texttt{user\_type} values across all review comments to determine reviewer composition. We count the number of ``User'' (human) and ``Bot'' (CRA) reviewers and assign one of five categories: \textit{CRAs} (only bots), \textit{User} (only humans), \textit{Mixed(CRAs)} (bots outnumber humans), \textit{Mixed(Human)} (humans outnumber bots), or \textit{Mixed} (equal counts). We also determine the review condition by examining the \texttt{state} attribute across all comments for each PR. We arrange PRs by their most definitive review status: \texttt{Approved} takes precedence, followed by \texttt{Dismissed}, then \texttt{Changes Requested}, with \texttt{Commented} as the default. For example, a PR with two \texttt{Commented} and one \texttt{Approved} review is classified as \texttt{Approved}.

We classify each PR into three outcomes using \texttt{state} and \texttt{merged\_at} attributes: \texttt{Merged} (state is ``closed'' and merged\_at is not null), \texttt{Closed} (state is ``closed'' but merged\_at is null, indicating abandonment), or \texttt{Stalled} (state is ``open'').

To answer RQ1, we examine PR outcomes across all review conditions. Only the \texttt{Commented} condition contains PRs reviewed solely by CRAs (281 PRs out of 2,456 total \texttt{Commented} condition PRs). Other conditions contain only mixed or human-only reviewers, showing that single CRA reviewers cannot independently approve, dismiss, or request changes. Since \texttt{Commented} is a neutral state where CRAs provide feedback without explicit decisions and the highest percentage (2,456 PRs out of 3,109), we use it as the basis for comparing human-only versus CRA-only effectiveness. We filter PRs with \texttt{review\_condition} equal to \texttt{Commented}, construct a contingency table with reviewer types as rows and PR outcomes as columns, and perform a chi-squared test of independence to validate statistical significance.



To answer RQ2, we analyze PRs reviewed solely by CRAs with \textit{Closed} outcomes to understand why CRA-only reviews failed to merge. We adopt the Signal-to-Noise Ratio framework from the blog~\cite{JetXu_2025} to evaluate review quality. We define two keyword tiers: \textbf{Tier 1 (Critical Signal)} includes runtime errors, crashes, compilation failures, API-breaking changes, and security vulnerabilities; \textbf{Tier 2 (Important Signal)} includes architectural problems, performance issues, and maintainability concerns. For each closed CRA-only PR, we count comments containing these keywords and classify them with a qualitative analysis technique called open coding~\cite{saldana2021coding}, achieving Cohen's Kappa coefficient of 0.75 for inter-rater agreement. We compute the signal-to-noise ratio as:

\begin{equation}
\text{Signal Ratio} = \frac{\text{Tier 1 Count} + \text{Tier 2 Count}}{\text{Total Comments}}
\end{equation}

This ratio ranges from 0.0 (no signal) to 1.0 (all actionable feedback). We categorize PRs into four signal ranges: 0-30\% (predominantly noisy, less than one-third actionable), 31-59\% (more noise than signal), 60-79\% (more signal than noise), and 80-100\% (predominantly actionable, high-quality feedback). These thresholds divide PRs based on whether signal or noise dominates their feedback content, allowing us to determine if noisy comments cause PR abandonment. Our implementation is publicly available~\cite{msr2026:challenge:CRA}.

\section{Analysis and Findings}
\subsection{Merge Rate Differences Between Human-Only and CRA-Only Reviews}

We begin by examining the distribution of review conditions across all 3,109 PRs. We find that \texttt{Commented} contains the majority, with 2,456 PRs (79.0\%), followed by \texttt{Changes Requested} with 492 PRs (15.8\%), \texttt{Approved} with 129 PRs (4.1\%), and \texttt{Dismissed} with only 32 PRs (1.0\%). Critically, PRs reviewed solely by CRAs appear exclusively in the \texttt{Commented} condition, as CRAs provide feedback without making explicit merge decisions.

We analyze the 2,456 PRs with \texttt{Commented} review condition to understand how reviewer composition affects outcomes. Table~\ref{tab:commented_outcomes} presents the distribution across five reviewer types: \textit{CRA}, \textit{User}, \textit{Mixed(CRA)}, \textit{Mixed(Human)}, and \textit{Mixed}. PRs reviewed solely by CRAs achieve 127 merged outcomes from 281 total PRs, yielding a 45.20\% merge rate. In contrast, human-only reviewed PRs merge 804 out of 1,176 PRs, achieving 68.37\%. This represents a 23.17 percentage point difference in merge success.

\begin{table}[h]
\centering
\small
\caption{PR Outcomes by Reviewer Type}
\label{tab:commented_outcomes}
\begin{tabular}{lrrrr}
\hline
\textbf{Reviewer Type} & \textbf{Total} & \textbf{Merged} & \textbf{Closed} & \textbf{Stalled} \\
\hline
CRA & 281 & 127 (45.20\%) & 98 (34.88\%) & 56 (19.93\%) \\
Mixed(CRA) & 117 & 74 (63.25\%) & 17 (14.53\%) & 26 (22.22\%) \\
Mixed & 604 & 369 (61.09\%) & 128 (21.19\%) & 107 (17.72\%) \\
Mixed(Human) & 278 & 189 (67.99\%) & 41 (14.75\%) & 48 (17.27\%) \\
User & 1,176 & 804 (68.37\%) & 254 (21.60\%) & 118 (10.03\%) \\
\hline
\end{tabular}
\end{table}

The table reveals significant differences in abandonment patterns. CRA-only reviewed PRs show a 34.88\% closure rate (98 out of 281 PRs), substantially exceeding the 21.60\% for human-only reviews (254 out of 1,176 PRs). The mixed reviewer categories demonstrate intermediate outcomes: human-dominated reviews achieve 67.99\%, CRA-dominated reviews show 63.25\%, and balanced mixed reviews achieve 61.09\%. This pattern suggests that human involvement substantially improves merge outcomes, even when CRAs are present. And the noticeable part is human-only user merged rate 23.17 percentage points higher than the CRA-only.

To validate statistical significance, we perform a chi-squared test of independence between reviewer type and PR outcome. The test yields $\chi^2 = 83.0319$ with 8 degrees of freedom and  $p < 0.001$, demonstrating a statistically significant association between reviewer type and merge outcomes.

\noindent 
\fbox{%
\begin{minipage}{0.97\linewidth} \textbf{Answer to RQ1:} CRA-only reviewed PRs achieve a 45.20\% merge rate, 23.17 percentage points lower than human-only PRs (68.37\%), with significantly higher abandonment (34.88\% vs 21.60\%). This difference is statistically significant ($\chi^2 = 83.0319$, $p < 0.001$). 
\end{minipage} }

\subsection{Signal-to-Noise Ratio in CRAs Comments}

\begin{figure}[htbp] 
 \centering 
 \includegraphics[width=\linewidth]{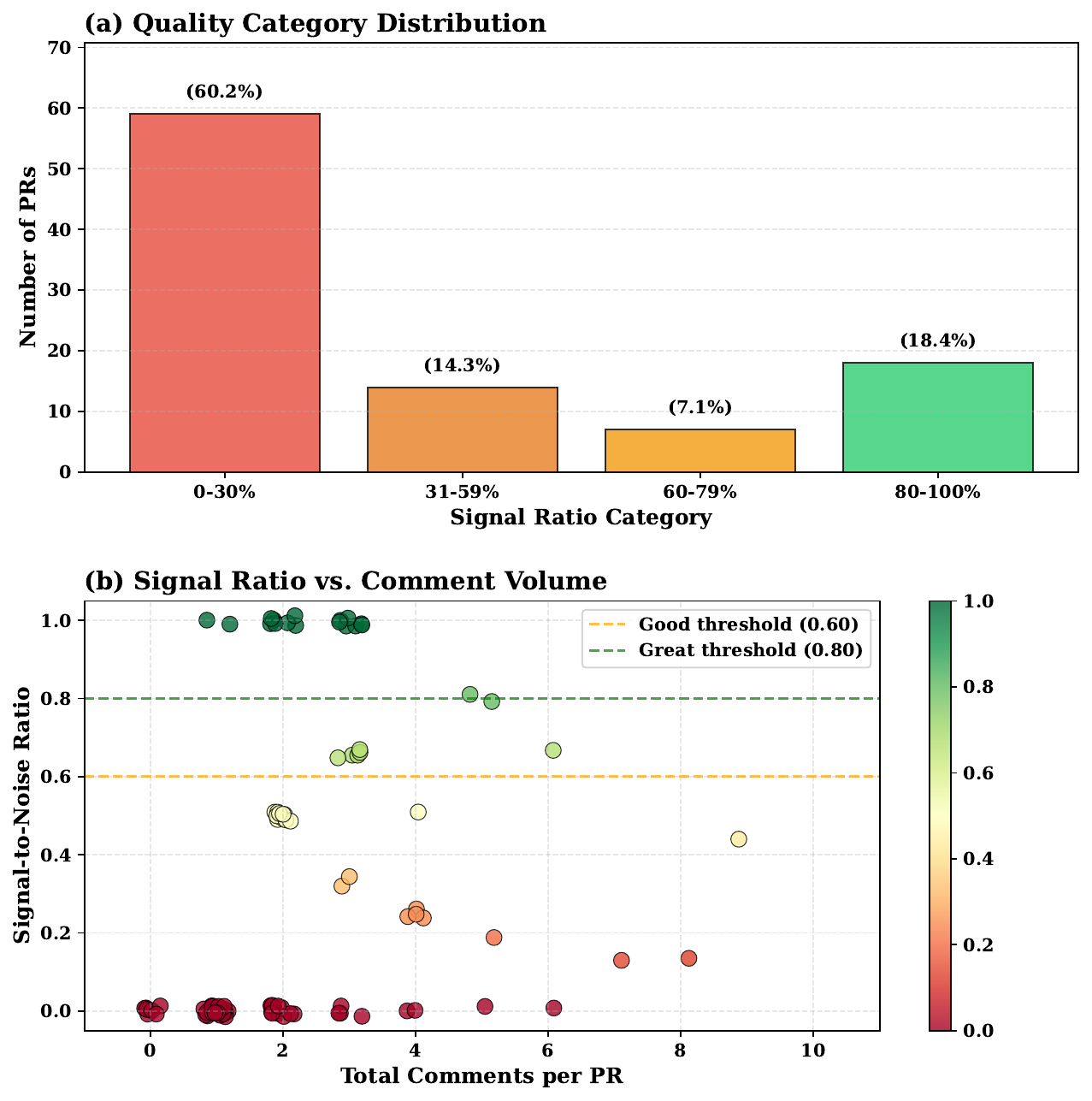} 
 \caption{Signal-to-noise ratio distribution across closed CRA-only PRs} 
 \label{fig:signal_distribution} 
 \vspace{-6.5pt}
\end{figure}

RQ1 revealed that CRA-only reviewed PRs experience a substantially higher abandonment rate (34.88\%) compared to other reviewer types in Table~\ref{tab:commented_outcomes}. To understand why CRA-only reviewed PRs show lower merge rates, we analyze the signal-to-noise ratio of the 98 closed CRA-only reviewed PRs. Figure~\ref{fig:signal_distribution}(a) presents the distribution across four signal ratio categories. The data reveals a heavily skewed pattern: 59 PRs (60.2\%) fall into the 0-30\% signal range, indicating more than half contain predominantly noisy comments. Additionally, 14 PRs (14.3\%) achieve 31-59\%, 7 PRs (7.1\%) reach 60-79\%, while 18 PRs (18.4\%) achieve 80-100\% signal ratio. This overwhelming concentration in the lowest category suggests that low signal-to-noise ratio is a primary factor in PR abandonment.

Figure~\ref{fig:signal_distribution}(b) shows the relationship between signal ratio and comment volume. PRs with great signal ratios ($\geq$0.80, dark green) cluster in the 1 to 3 comment range, achieving ratios near 1.0, indicating high-quality CRA reviews provide concise feedback. PRs with good signal ratios (0.60-0.79, light green) scatter between 2 to 6 comments. In contrast, PRs with poor signal ratios ($<$0.30, dark red) concentrate heavily at the bottom across 1 to 8 comments, exhibiting ratios near 0.0. The middle-quality PRs (0.30-0.59, yellow-orange) scatter throughout with ratios between 0.20-0.50. This demonstrates that comment volume alone does not determine quality; the proportion of actionable feedback is critical.

We investigate the 98 closed PRs to identify which CRAs contribute to the noise problem. We find 13 unique CRAs, with github-advanced-security[bot] having the highest count (36 PRs), followed by Copilot (24 PRs) and ellipsis-dev[bot] (11 PRs). However, PR count does not correlate with quality. Copilot achieves only a 19.79\% average signal ratio, while github-advanced-security[bot] achieves 27.62\%. Multiple CRAs show very low ratios: codefactor-io[bot] at 0.00\%, aikido-pr-checks[bot] at 20.00\%, semgrep-code-assistant-ws[bot] at 25.00\%. Only semgrep-code-getsentry[bot] achieves 100.00\%, though it reviewed just 1 PR. Among CRAs with substantial counts, entelligence-ai-pr-reviews[bot] shows the highest ratio at 52.29\% (7 PRs), followed by cursor[bot] at 43.40\% (5 PRs). Critically, 12 of 13 CRAs (92.31\%) exhibit average signal ratios below 60\%.

\noindent 
\fbox{%
\begin{minipage}{0.97\linewidth} \textbf{Answer to RQ2:} Among 98 closed CRA-only PRs, 60.2\% show 0--30\% signal ratios, and 12 of 13 CRAs exhibit ratios below 60\%, confirming that low signal-to-noise ratio substantially contributes to PR abandonment. 
\end{minipage} }

\section{Threats to Validity}

\textbf{Construct Validity:} Our keyword-based signal-to-noise classification may miss actionable feedback without keywords or incorrectly label keyword-containing but irrelevant comments. We manually validated keyword-matched comments to mitigate this. Our priority-based review state classification was cross-checked against GitHub labels for consistency.

\textbf{Conclusion Validity:} Manual validation can introduce bias or misclassification. To mitigate this, two researchers independently classified comments, discussed disagreements, and iteratively refined keyword lists. While our chi-squared test shows statistically significant associations between reviewer type and PR outcomes, correlation does not imply causation.

\textbf{External Validity:} Our findings are based on 3,109 PRs from the AIDev dataset on GitHub repositories with AI-generated code, which may not generalize to proprietary repositories, other platforms, or projects without CRAs. However, the dataset spans diverse domains, sizes, and maturity levels, providing a strong basis for understanding CRA behavior in open-source environments.

\vspace{-4pt}
\section{Related Work}

Studies have examined AI coding agents in PR workflows. Watanabe et al.\ showed that agentic PRs are used for maintenance tasks like refactoring, documentation, and test generation~\cite{Watanabe_2025_AgenticPR}. Horikawa et al.\ reported that AI agents perform refactorings, including renaming and type-related edits~\cite{Horikawa_2025_AgenticRefactoring}. Twist found that agents import libraries but rarely add new dependencies~\cite{Twist_2025_LibraryUsage}. These studies focus on what PRs change, not how review comments affect outcomes.

Researchers have studied CRAs in open-source software. Wessel et al.\ found that agents reduce manual work and standardize feedback, but introduce noise and harm communication when poorly aligned~\cite{Wessel_2020_BotsSurvey}. Recent work examines LLM-based review tools in industry. Ramesh et al.\ reported that developers found an LLM assistant helpful but relied on human judgment for decisions~\cite{Ramesh_2025_Ericsson}. Vijayvergiya et al.\ deployed AutoCommenter at Google, showing these tools can work at scale~\cite{Vijayvergiya_2024_AutoCommenter}. These studies examine tool adoption and developer views rather than PR outcomes.

Other studies investigate whether LLMs correctly understand code review content. Lin et al.\ proposed CodeReviewQA, showing that many models struggle with identifying review intent and required changes~\cite{Lin_2025_CodeReviewQA}. This indicates AI-generated comments may suffer from limited precision. Beyond review generation, researchers examined how context shapes workflows. Chatlatanagulchai et al.\ found developers often provide build instructions but less frequently document security and policy constraints~\cite{Chatlatanagulchai_2025_AgentREADMEs}. TiCoder shows that structured interaction reduces cognitive load and improves evaluation of AI-generated code~\cite{Fakhoury_2024_TiCoder}.

Prior work has explored agent-authored PRs~\cite{Watanabe_2025_AgenticPR}~\cite{Horikawa_2025_AgenticRefactoring}~\cite{Twist_2025_LibraryUsage}, traditional CRAs~\cite{Wessel_2020_BotsSurvey}, LLM-based review tools~\cite{Ramesh_2025_Ericsson}~\cite{Vijayvergiya_2024_AutoCommenter}~\cite{Lin_2025_CodeReviewQA}, and context in workflows~\cite{Chatlatanagulchai_2025_AgentREADMEs}~\cite{Fakhoury_2024_TiCoder}. However, these studies focus on authoring behavior or perception rather than how reviewer composition and comment quality relate to merge and abandonment outcomes.

\section{Discussion}

 \textbf{The Need for Human Code Reviewers:} Our analysis reveals two critical takeaways. First, CRAs cannot effectively replace human reviewers: while industry reports claim 80\% of PRs need no human comments~\cite{QodoReport_2025}~\cite{GreptileReport_2025}, CRA-only reviews achieve only 45.20\% merge rates versus 68.37\% for human-only reviews, with 34.88\% abandonment rates. Second, most CRAs generate predominantly noisy feedback: signal-to-noise analysis shows 60.2\% of closed CRA-only PRs fall into the 0-30\% signal range, and 92.31\% of CRAs perform below 60\% signal quality. This noise burden overwhelms developers, forcing them to sift through irrelevant feedback, increasing cognitive load without providing clear improvement paths.

\textbf{Effective Use Cases for Code Review Agents:} For developers, we provide actionable strategies by identifying concrete use cases for code review agents. Developers should configure CRAs for narrow, specific checks such as security vulnerabilities or style violations rather than general-purpose review, because specialized checks reduce false positives and improve feedback relevance. Specialized CRAs like semgrep-code-getsentry[bot] demonstrate higher precision in the AIDev dataset. Enforce human approval before merging, as human-dominated reviews achieve 67.99\% merge rates by providing contextual judgment that CRAs lack. Establish workflows where CRA comments trigger human review rather than direct developer responses, combining automation efficiency with human judgment on design and architectural decisions.

\section{Conclusion}

This paper investigates the effectiveness of CRA code reviews in Agentic PR workflows. While industry reports claim that 80\% of PRs receive no human comments when AI review is enabled, our analysis of 3,109 PRs reveals a contrasting reality. CRA-only reviewed PRs achieve a 45.20\% merge rate, significantly lower than the 68.37\% rate for human-only reviewed PRs, with substantially higher abandonment rates (34.88\% vs 21.60\%). Investigation of the 98 closed CRA-only PRs reveals that 60.2\% fall into the 0-30\% signal range, and 12 of 13 unique CRAs (92.31\%) exhibit average signal ratios below 60\%. This demonstrates that current CRAs frequently generate noisy feedback with minimal actionable content, which increases abandonment risk and burdens developers with low-value comments. The mixed reviewer results show that human involvement substantially improves outcomes.

As autonomous coding agents generate code at an unprecedented scale, our findings reveal an urgent need of extend this work in three directions. First, develop metrics that quantify actionable signals in automated review comments beyond merge outcomes. Second, analyze reviewer effort by examining comment volume, requested changes, and integration feedback to design CRAs that reduce review burden. Third, investigate decision models that determine when automated reviews require human involvement based on review outcomes and past agent performance. Pursuing these directions can enable empirically grounded review systems that improve PR outcomes while mitigating the risk of abandonment.

\bibliographystyle{ACM-Reference-Format}
\bibliography{references}

\end{document}